\documentclass{IEEEtran}
\usepackage{graphicx}
\usepackage{multirow}
\usepackage{float}
\usepackage[T1]{fontenc}
\usepackage[utf8]{inputenc}
\usepackage{tabularx}
\usepackage{algorithm}
\usepackage{algpseudocode}
\usepackage{amsmath,amssymb,amsfonts}
\usepackage{lettrine}
\usepackage{cite}
\usepackage[stable]{footmisc}
\usepackage{xcolor}
\usepackage{comment}

\title{Implications of the Reciprocity Theorem for Reconfigurable Intelligent Surfaces}

\author{Uday~K~Khankhoje,~\IEEEmembership{Senior~Member,~IEEE} and 
Debidas~Kundu,~\IEEEmembership{Senior~Member,~IEEE}%
\thanks{This work is generously supported by (i) Ministry of Electronics and Information Technology (MeitY), Government of India under a grant ``6G : Sub-THz Wireless communication with Intelligent Reflecting Surfaces(IRS)" number R23011/3/2022-CC\&BT-MeitY, and (ii) Qualcomm 6G University Research India Program under a grant: ``Intelligent approaches to designing and building Intelligent Reflecting Surfaces for 6G applications.”}%
\thanks{\textit{Corresponding author: Uday K Khankhoje}.}%
\thanks{Uday K Khankhoje is with the Department of Electrical Engineering, Indian Institute of Technology Madras, India, e-mail: uday@ee.iitm.ac.in.}%
\thanks{Debidas Kundu is with the Department of Electrical Engineering, Indian Institute of Technology Delhi, India, e-mail: debidask@ee.iitd.ac.in.}%
}

\begin{document}
\maketitle

\begin{abstract}
Reciprocity between a transmitter and receiver is a foundational requirement in wireless communications. A few recent works have suggested that reciprocity is broken under reflection by reconfigurable intelligent surfaces (RIS) when the reflection phase becomes incident angle dependent. In this work, we rigorously show that these claims are based on the use of idealized reflection coefficients that ignore mutual coupling between heterogeneous unit cells, surface-truncation effects, and structural scattering contributions from the RIS. Full-wave electromagnetic simulations of transmit/receive antennas and a finite-size RIS implemented via a particular unit cell design are performed to quantitatively demonstrate that reciprocity holds even in the presence of incident-angle dependent reflection phases. To show this, we calculate two-port antenna scattering parameters and evaluate the electromagnetic reciprocity integral to support our claims.
\end{abstract}


\section{Introduction}
\lettrine[lines=2]{R}econfigurable intelligent surfaces (RIS) have been proposed as key ingredients for enhancing coverage, especially around obstacles, in upcoming 6G wireless communications \cite{6g_a_research_direction}. This has led to a flurry of research works, both in the antennas and propagation community \cite{zhang_dual_pol_2022, Sarris_propag_modeling_RIS}, as well as in the wireless communications community \cite{Di_Renzo_Smart_Radio_Environments_2020}, with each community working on different aspects of design, hardware implementation, and analysis of RIS and associated communications problems. 

An RIS is typically an electromagnetic metasurface composed of sub-wavelength-size repeating unit cells. Each unit cell is equipped with a tunable element, such as a PIN or a varactor diode, such that the reflection phase can be changed by electronically modulating the diode state \cite{zhang_dual_pol_2022, Emara_reconfigurable_metasurface_varactor_2024}. By changing the phases across the RIS appropriately, an incoming electromagnetic wave can be reflected towards a desired direction. It is critical that this process be reciprocal in order to establish bi-direction communication between the two parties, a foundational assumption in wireless communications. Here, ``reciprocal'' qualitatively means that under time-reversal a wave will retrace its path from the receiver to the transmitter, thereby establishing the bidirectional nature of communication (we reserve the quantitative elaboration on reciprocity for later). 

A few recent works involving RIS have suggested that metasurfaces are not angle reciprocal when the unit cell reflection phase depends on the incidence angle \cite{chen_angle-dependent_2020, Yue_reciprocity_tvt_2023, 10360391}. The reasoning provided is that since the incident and desired reflection angles are different, so are the (incident angle dependent) reflection phases in each direction of propagation from the RIS. Thus, by invoking Fermat's principle of least time, i.e.~using geometric optics and by taking propagation and reflection phases into account \cite{yu_light_propag_2011}, the time-reversed wave from the receiver need not travel in the direction of the reversed original incident wave, thus leading to a loss of reciprocity in the system. 
Obviously, such a consequence has profound implications on wireless communications and requires proper redressal. This observation has spawned numerous works \cite{cui_3_bit_angle_insensitive_2022, liang_low_angular_sensitivity_2024, zhao_angular_stability_for_precise_wave_manipulation_2025} that aim to design unit cells whose reflection phases are (ideally) independent of the incidence angle. 

We show in our work that while the design of unit cells with incidence angle-independent reflection coefficients has many performance-related advantages, concerns of reciprocity breaking are largely unfounded. The electromagnetic reciprocity theorem itself guarantees angle reciprocity when the reflecting structure is characterized as a linear, inhomogeneous medium describable by symmetric permittivity and permeability tensors \cite[Ch.~30]{chewemt}. So, where does the argument of reciprocity violation fail? The key flaw is the study of finite-size RIS structures using reflection coefficients obtained from electromagnetic simulations of unit cells using periodic-boundary conditions, which we term as ``ideal'' reflection coefficients. Applying the reflection coefficients so obtained ignores mutual coupling effects between dissimilar unit cells (dissimilar due to different diode states), phase perturbations that occur on account of surface truncation, and structural scattering contributions from the RIS \cite[Eq.~2-130]{balanis2016antenna}. This, in turn, leads to an erroneous prediction of reciprocity violation. We show via full-wave simulations how  reciprocity is indeed maintained when  non-idealities are taken into account.

\section{How Ideal Reflection Coefficients Suggest Reciprocity Violation}\label{secideal}

Assume plane wave propagation as described by the electric field: $\vec{E}(\vec{r},t) = \vec{E}_0 \exp(j(\omega t - \vec{k}\cdot\vec{r}))$ ($\omega$: angular frequency, $\vec{k}$: wavevector in the $x-z$ plane with magnitude $k$), making an angle $\theta$ as defined from the $+z$ axis (in the clockwise sense), impinging on an electromagnetic surface in the $x-y$ plane. For simplicity, let the surface consist of $n$ elements arrayed in the $x-$direction, each spaced $d$ apart and imparting a (possibly incident angle dependent) phase $\psi_i$ on reflection. The inter-element phase relationship at the array elements due to path length differences of the incoming wave is defined in terms of the array steering vector,  $a(\theta)$, (hereon we omit the time-dependence in $\vec{E}$): 
\begin{equation}
a(\theta) = \left[1,\,e^{jkd\sin\theta},\ldots,e^{jk(n-1)d\sin\theta}\right]^T.
\end{equation}

\begin{figure}[h!]
   \centering 
    \includegraphics[clip, trim = 0cm 0.8cm 0cm 0cm, width=0.7\columnwidth]{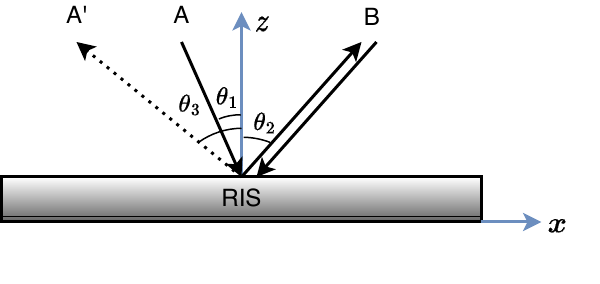}
    \caption{Geometry of ray propagation for the forward ($A \rightarrow B$) and reverse ($B \rightarrow A'$) directions after reflection from the RIS.}
    \label{fig2}
\end{figure}

Now consider a wave incident at $\theta_1$ and wave reflected in the $\theta_2$ direction (in the specular quadrant) as shown in Fig.~\ref{fig2} (note: $\theta_1,\theta_2 > 0$). The corresponding wavefront of the reflected wave has the following relation (up to phase constants) due to reflections from the $n$ elements: 
\begin{equation}\label{yeq}
y = a(\theta_2) \cdot \Psi \cdot a(-\theta_1),
\end{equation}
where $\cdot$ denotes element wise multiplication and $\Psi$ is a vector consisting of element-wise reflection coefficients, $R_ie^{j\psi_i}$. In particular: 
\begin{equation}
y_i = e^{jk(i-1)d\sin\theta_2} \, R_ie^{j\psi_i} \, e^{-jk(i-1)d\sin\theta_1}.
\label{phasefront}
\end{equation}

\subsection{RIS with continuous phase shifts}

We infer from \eqref{phasefront} that for an RIS with continuous phase shifts, the relation for constructive interference along the $\theta_2$ direction, i.e.~the path $A\rightarrow\text{RIS}\rightarrow B$ in Fig.~\ref{fig2}, all the $y_i$'s are in phase when the reflection phases from each element, $\psi_i$, satisfy (up to a constant):
\begin{equation}\label{psi1to2}
\psi_i = (i-1)\underset{\Delta\psi}{\underbrace{kd(\sin\theta_1 - \sin\theta_2)}}.
\end{equation}
These reflection coefficient phases are realized by the specific geometry of the unit cell and the tunable elements, such as varactor diodes, embedded therein. For sake of argument, assume that the values of $\psi_i$ required above are realized by appropriately tuning the diode voltages, and are ``frozen'' for the next step.

In the next step, consider the ``reciprocal'' case, where the wave is incident from the reverse direction, i.e.~$\theta_2$ (with $\theta_2 \neq \theta_1$), and in general goes to some other direction $\theta_3$ ($>0$ and possibly different from $\theta_1$, but in the same quadrant), as seen in Fig.~\ref{fig2}. By analogy with the above analysis, the corresponding reflected wavefront has the following phases (up to a constant) due to reflections from the $n$ elements: 
\begin{equation}\label{zeq}
z = a(-\theta_3) \cdot \Psi' \cdot a(\theta_2),
\end{equation}
where $\Psi'$ is a vector containing the modified reflection coefficients, different from $\Psi$ since the incidence angle is different. These modified coefficients are represented as follows:
\begin{equation}\label{dpsid}
\psi_i' = (i-1) \Delta \psi' .
\end{equation}
This leads to the reflected wave's front having phases as:
\begin{equation}
\text{arg}(z_i) =(i-1)[kd(\sin\theta_2 - \sin\theta_3) + \Delta \psi']. \label{nonreci}
\end{equation}
For us to have constructive interference at $\theta_3$ in the above case, we require the following to be true (obtained after substituting \eqref{psi1to2} and \eqref{dpsid} into \eqref{nonreci} and some algebra):
\begin{equation}\label{recicont}
\sin\theta_3 = \sin\theta_1 + \frac{(\Delta \psi' - \Delta \psi) - 2m\pi}{kd}, \,\, m\in\mathbb{Z},
\end{equation}
where the last term ensures a real value for $\theta_3$ \footnote{A similar result is obtained in \cite[Eq.~4]{chen_angle-dependent_2020}.}.

There are two implications of this result that deserve note;
\begin{enumerate}
\item If the reflection phases are independent of the incidence angle, then $\Delta \psi' = \Delta \psi$, which automatically leads to $\theta_3 = \theta_1$, implying that reciprocity is maintained.
\item If however, the reflection phases depend on the incidence angle, then $\Delta \psi' \neq \Delta \psi$ in general and the return path beam need not be formed at $\theta_1$, implying that reciprocity has broken.
\end{enumerate}

\subsection{RIS with discrete phase shifts}
A natural question is whether the so-called reciprocity breaking also happens in the case of RIS realizations where only a few discrete states, as in the case of a PIN diode, are permissible. In the earlier case where arbitrary reflection phases were available for beamforming (e.g.~as per \eqref{psi1to2}), in the discrete case we show that there are no closed-form expressions for the beamforming direction in the ``reciprocal'' case (like in \eqref{recicont}).

To see this, we recall that forming a beam at a particular angular location corresponds to maximizing the array factor at that location. In our recent work \cite{sai}, we compute the optimal beamforming weights, $w\in\mathcal{B}$, where $\mathcal{B}$ is a finite set. These discrete weights correspond to the field-reflection coefficients for various diode states. We express the array factor, $G$, for beamforming from  $A\rightarrow \text{RIS}\rightarrow B$, as:
\begin{equation}\label{origaf}
G(-\theta_1,\theta_2) = \sum_{i=1}^n w_i \, \exp(j(i-1)kd[\sin\theta_2 -\sin\theta_1]).
\end{equation}
Then, the beamforming weights $w$ (i.e.~the reflection coefficients), are obtained by solving the optimization problem:
\begin{equation}\label{discretopt}
\max_{w} \quad |G(-\theta_1, \theta_2)|,\,\text{s.t.}~w\in\mathcal{B}.
\end{equation}
Assume we have solved this problem and obtained the optimal weights, say $\tilde{w}$. However, for beamforming in the reverse direction ($B\rightarrow \text{RIS}\rightarrow A$), since the weights are incidence-angle dependent, they would have transformed from $\tilde{w}$ to (possibly) different weights, $\tau$. Therefore, the corresponding array factor expression becomes:
\begin{equation}\label{revaf}
G(\theta_2, -\theta_1) = \sum_{i=1}^n \tau_i \exp(j(i-1)kd[\sin\theta_2 -\sin\theta_1]).
\end{equation}
There is no particular reason for \textit{this} $\tau$ to be the optimal set of weights for beamforming from $B\rightarrow \text{RIS}\rightarrow A$, i.e.~to maximize $|G(\theta_2, -\theta_1)|$. Hence, the weights $\tilde{w}$, which are optimal in the forward direction, are not necessarily optimal in the reverse direction, thereby leading to a seeming violation of reciprocity even in the case of an RIS with discrete weights.

Interestingly, if all the weight-phases are transformed by the same amount, say $\kappa$, i.e.~$\tau_i = \exp(j\kappa)\, \tilde{w}_i,\,\forall i$, it leads to the condition of optimal gain in the reverse direction, i.e.~$|G(\theta_2, -\theta_1)| = |G(-\theta_1, \theta_2)|$. For reciprocity to hold under the array factor approach, we would actually require $G(\theta_2, -\theta_1) =  G(-\theta_1, \theta_2)$ to be true\footnote{This can be shown by expressing S-parameters in terms of antenna circuit models and the scattering matrix of the RIS under plane-wave and far-field assumptions; see, e.g., Ch.~2 of \cite{balanis2016antenna}.}; however, this is not so in the special case considered here, since $G(\theta_2, -\theta_1) = \exp(j\kappa)\,G(-\theta_1, \theta_2)$.
\section{The Restoration of Reciprocity via Full-wave Calculations}\label{sec3}

The earlier section showed that under the array factor (AF) approach, the RIS is predicted to be non-reciprocal when unit cell reflection phases are incidence-angle dependent. However, this approach is an incomplete model of the scattered field: the full scattered field from an RIS consists of a structural scattering term and an antenna-mode term \cite[Eq.~2-130]{balanis2016antenna}, whereas the AF approach retains only the antenna-mode term while also ignoring inter-element mutual coupling. Ignoring these contributions leads to the erroneous prediction of non-reciprocity. The correct determination of reciprocity follows from the Lorentz reciprocity theorem, which guarantees that any structure free of nonlinear, active, or magneto-optic elements is reciprocal \cite{harrington1961time}.

We now examine the numerical implications of the reciprocity theorem by considering a particular design of a finite-size RIS. 

\begin{figure}[h!]
   \centering
   \includegraphics[width=1.00\columnwidth]{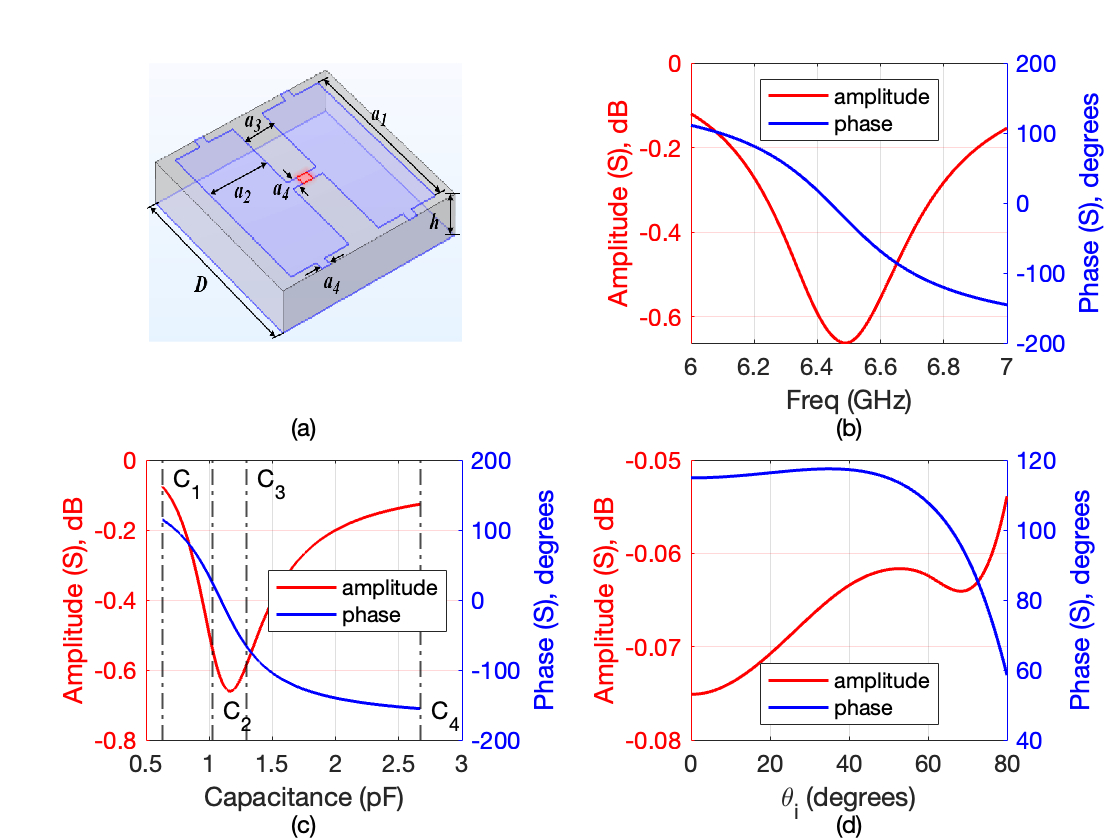}
    \caption{(a) Unit cell geometry and parameters (values are given in Appendix A), (b) $C=1.167$ pF,  $\theta_i=0^\circ$ (c) $f=6.5$ GHz, $\theta_i=0^\circ$, $C_{i}=\{ 0.63,\, 1.025,\,1.291,\, 2.67\}$ pF, (d)  $f=6.5$ GHz, $C=0.63$ pF.}
    \label{unitcellfig}
\end{figure}

\textit{Unit cell design:} For definiteness, we consider a unit cell similar to the one discussed in prior work \cite{chen_angle-dependent_2020}. This unit cell is loaded with a varactor diode and in order to implement 2-bit beam forming, we optimize its dimensions to obtain a $270^\circ$ phase swing in the reflection coefficient at 6.5 GHz, as shown in Fig.~\ref{unitcellfig} (refer to the Appendix for the unit cell parameters).

\begin{figure}[h!]
    \centering
    \includegraphics[width=0.65\columnwidth]{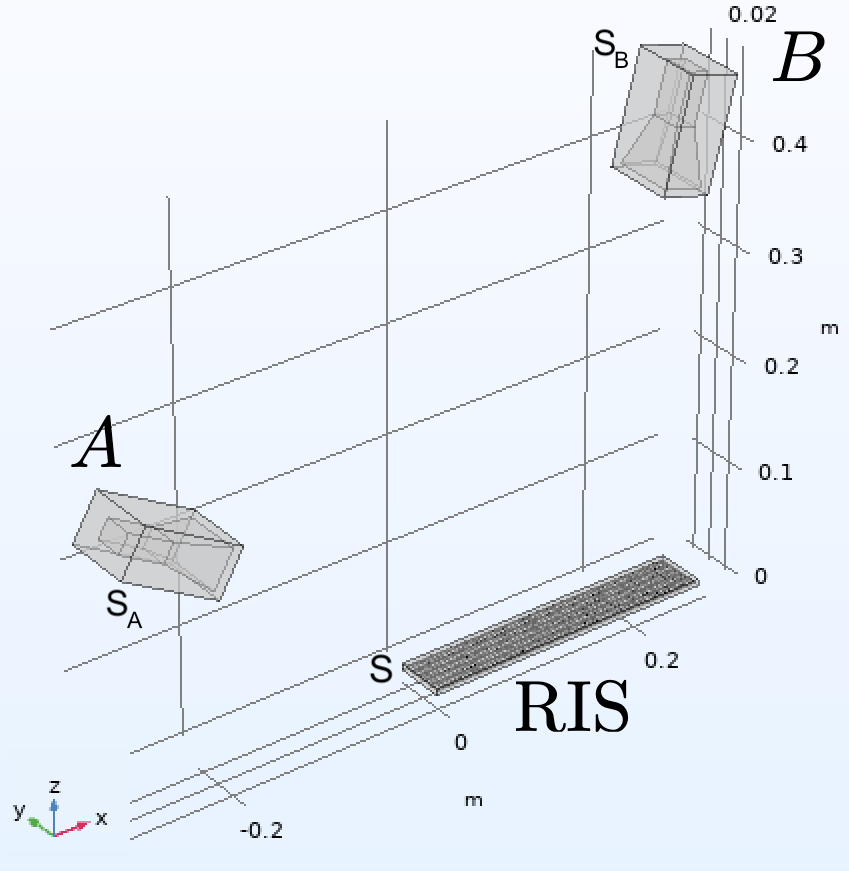}
    \caption{Simulation setup for numerical experiments, with horn antennas at positions $A$, $B$, and the $32\times 5$ RIS in the $xy$ plane. All three entities are enclosed in rectangular domains which couple their physics to each other via the finite element method -- boundary integral method.}
    \label{figfullsetup}
\end{figure}

\textit{Simulation setup: } Consider two horn antennas designed for single mode operation at 6.5 GHz, one kept at location $A$, generating fields $\vec{E}_1(\vec{r}),\vec{H}_1(\vec{r})$, and the other at location $B$, generating fields $\vec{E}_2(\vec{r}),\vec{H}_2(\vec{r})$. These sources are kept at a radial distance of $10\lambda$ and have angular coordinates $65^\circ \text{ and } -15^\circ$ (w.r.t.~$+z$ axis), respectively, with regards to the center of the $32\times 5$ sized RIS kept in the $xy$ plane, as shown in Fig.~\ref{figfullsetup}. Note from Fig.~\ref{unitcellfig}(d) that the reflection phases at $\theta_i=-65^{\circ}$ and $\theta_i=15^{\circ}$ are $15^{\circ}$ apart. We now use the optimal partitioning algorithm (OPA) method \cite{sai} to obtain optimal weights for beamforming from $A\rightarrow \text{RIS} \rightarrow B $. The RIS weights, i.e.~the varactor capacitances, are kept constant for all numerical experiments. Full-wave electromagnetic simulations are performed in Comsol Multiphysics, where we use the finite element -- boundary integral (FE-BI) method. Further technical details of the simulation are mentioned in the Appendix.

\begin{figure}[htbp]
    \centering
    \includegraphics[width=0.85\columnwidth]{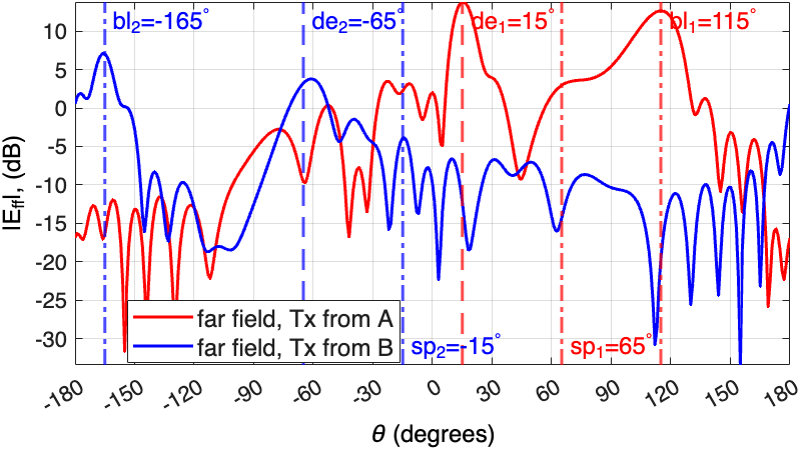}
    \caption{Far field electric fields in the two cases of the transmitter being at positions $A$ (red) and $B$ (blue), respectively, with the angular coordinate $\theta$ measured clockwise w.r.t.~$+z$ axis. The vertical lines with labels `de' and `bl' denote desired reflection and back lobe directions, respectively, for each case.}
    \label{ff-param}
\end{figure}

\textit{Qualitative appreciation of reciprocity:} To get a visual representation of the far fields corresponding to illuminations from locations $A$ and $B$ (in serial order), we refer to Fig.~\ref{ff-param}. In the intended case of illumination from $A$ (i.e.~from $-65^{\circ}$), we observe a far field peak near the desired direction at $15^{\circ}$ due to the chosen coding pattern. Additional peaks arise at the specular direction ($65^{\circ}$) and in the forward transmission direction ($115^{\circ}$) due to the finite nature of the RIS. For considering reciprocal communication we now examine illumination from $B$ (i.e.~from $15^{\circ}$) and observe a visible peak near the intended direction (i.e.~$-65^{\circ}$), thereby establishing reciprocity in a qualitative sense. In this case as well, we observe additional peaks in the specular ($-15^{\circ}$) and forward transmission  ($-165^{\circ}$) directions. Note: these additional peaks are not captured by the array factor approach.

\textit{First experiment -- Scattering-parameters:} In the first quantitative experiment, reciprocity is proved via the use of scattering parameters. In this, the antennas at $A$ and $B$ are made the transmitter and receiver, respectively, and the $S$-parameters, in particular $S_{21}$, are computed. Next, with the configuration of the RIS unchanged, the antennas at $A$ and $B$ are made the receiver and transmitter, respectively, and $S_{12}$ is computed. The classical formulation of the reciprocity theorem applied to two sources gives the equality of reaction integrals \cite[Eq.~3-38]{harrington1961time}, $\langle a,b \rangle = \langle b,a \rangle$, which for a two-port network translates directly to $S_{12} = S_{21}$.

\begin{figure}[htbp]
    \centering
    \includegraphics[width=0.85\columnwidth]{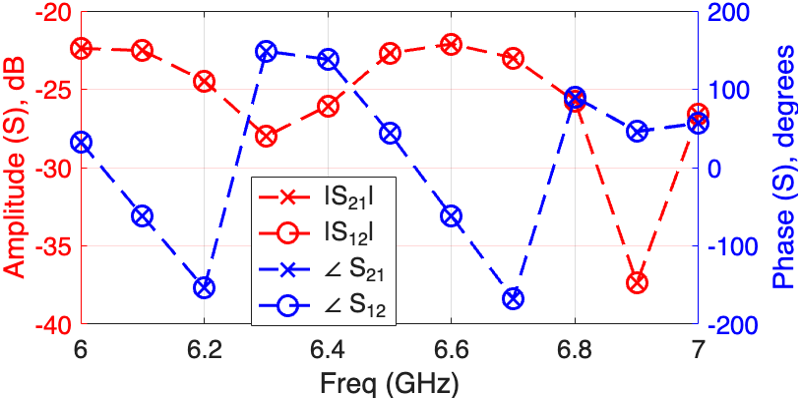}
    \caption{$S_{12}$ and $S_{21}$-parameters for the first numerical experiment.}
    \label{s-param}
\end{figure}

We compute the S-parameters in Comsol Multiphysics and report the results in Fig.~\ref{s-param}, observing that  $S_{12} = S_{21}$ to a very high degree of numerical accuracy, in particular we find that the average disagreement between $S_{12}$ and $S_{21}$ is less than 0.0032 dB, thereby establishing reciprocity in a straightforward way. We further note that the result holds for a wide range of frequencies, i.e.~within the frequency range when the structure is operating like an RIS, as well as outside the frequency range of the resonance, i.e.~when the structure is operating like a homogeneous reflecting metal sheet, thereby demonstrating the robustness of the result.

\textit{Second experiment -- reciprocity theorem: } In the second experiment, reciprocity is directly proved via application of the electromagnetic reciprocity theorem. A volume just enclosing the RIS does not contain any impressed sources, and the applicable source-free version of the reciprocity theorem \cite[Eq.~3-34]{harrington1961time} gives this surface integral identity:
\begin{equation}\label{recieqn}
\Psi = \oint_S \left(\vec{E}_1\times \vec{H}_2 - \vec{E}_2 \times \vec{H}_1 \right) \cdot  d\vec{s} = 0,
\end{equation}
where $S$ is a surface enclosing the RIS as seen in Fig.~\ref{figfullsetup}, and the subscripts $\{1,2\}$ on the electric $(E)$ and magnetic $(H)$ denote the fields that are generated when transmitter is at positions $A$ and $B$, respectively. 

\begin{figure}[htbp]
    \centering
    \includegraphics[width=0.85\columnwidth]{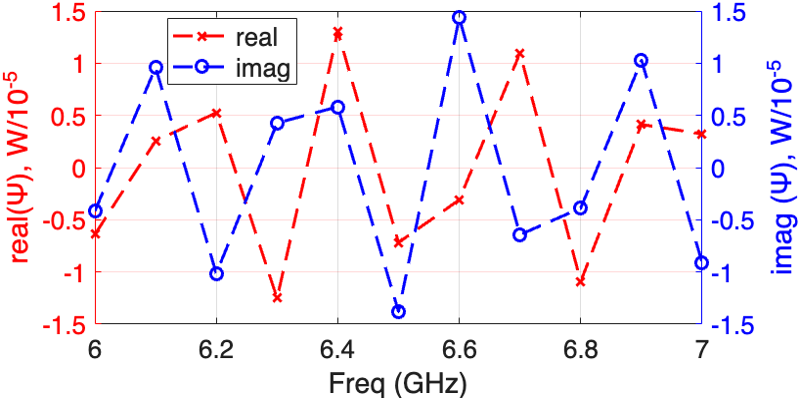}
    \caption{Real / imag parts of $\Psi$ (Eq.~\eqref{recieqn}) for the second numerical experiment.}
    \label{r-param}
\end{figure}

As observed in the case of the first experiment, here also the above equation \eqref{recieqn} is found to be satisfied to a high degree of numerical accuracy over a wide range of frequencies. The quantity $\Psi$, from \eqref{recieqn} is computed in Comsol Multiphysics and plotted in Fig.~\ref{r-param}; in particular we find the average value of $|\Psi|$ to be less than $3.7\times10^{-6}$ W over the frequency range considered. As a consequence of both the numerical experiments, reciprocity in the presence of the RIS is indisputably established. 

\textit{Actual v/s ideal reflection coefficients:} The analysis shown in Sec.~\ref{secideal} implicitly relied on the use of the ideal reflection coefficients. A common practice in the RIS literature is to simulate a unit cell of the RIS with periodic boundary conditions, i.e.~the \textit{same} unit cell is repeated periodically in space. Full wave simulations are performed for all discrete states and the reflection coefficients are stored in each case. Next, given the directions of the incident and reflected beams,  beam forming algorithms are applied to a finite-size RIS to determine the discrete states of each of the cells. The key assumption made here is that the reflection coefficient of a cell is not altered by its heterogeneous neighbors. As an example, in a 2-bit RIS the discrete set $\mathcal{B}$ in \eqref{discretopt} has cardinality 4, which implicitly assumes that the reflection coefficients for each unit cell in the finite RIS can take only four possible (ideal) values. 

\begin{figure}[h!]
    \centering
\includegraphics[clip, trim = 0.3cm 0.8cm 0.9cm 1.2cm, width=1.0\columnwidth]{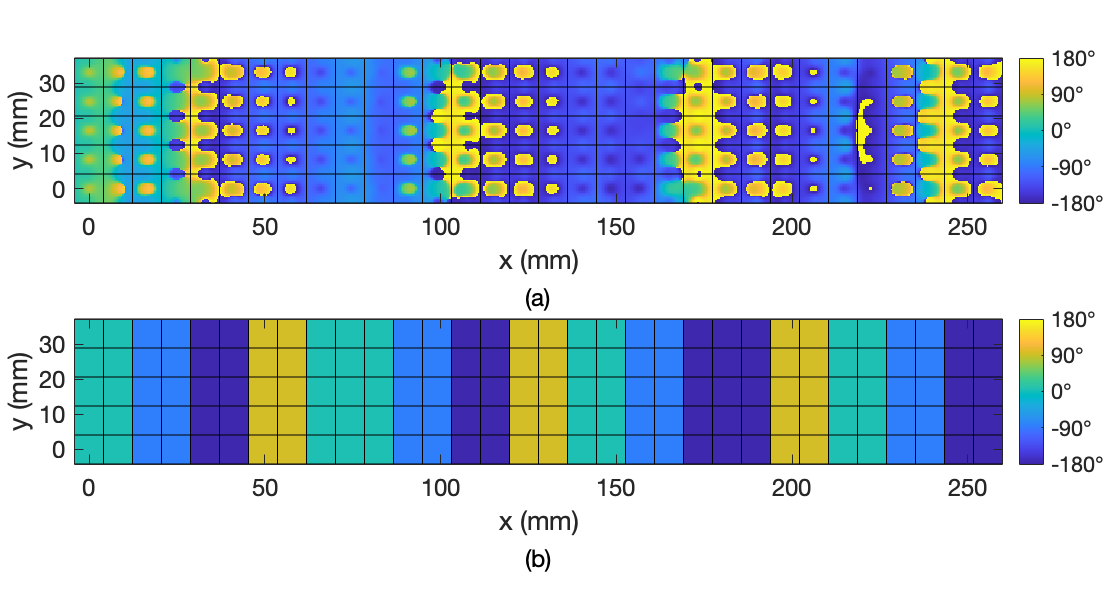}
    \caption{Phase of (a) actual, and (b) ideal reflection coefficients across the RIS. These phases are matched at the left-most column, middle row unit cell.}
    \label{refcoef}
\end{figure}

However, it is expected that due to inter-cell mutual coupling effects, the reflection coefficients can vary depending on the local neighborhood of a given unit cell. We see this numerically in Fig.~\ref{refcoef}, where we plot the actual and ideal reflection coefficients in the beam forming configuration of $\text{Tx}(A)\rightarrow \text{RIS} \rightarrow \text{Rx}(B) $. Thus the actual beam forming problem is considerably more complex than the one suggested in \eqref{discretopt}, since the set $\mathcal{B}$ itself depends upon the spatial arrangements of the unit cells. An exact solution to this problem involves optimization on a finite RIS, which is a particularly intense computational undertaking and therefore not practical. That said, the reader may refer to recent work \cite{Abrardo2024} on electromagnetic-theory compliant multi-port models of RIS and antennas that capture mutual coupling, finite surface effects (e.g.~specular scattering), and predict symmetric scattering coefficients. 

We remark that the notion of an ``actual'' reflection coefficient does not convey the same physical meaning as an ``ideal'' coefficient for the reason that the former is a local quantity defined as the ratio of the numerical $H_y$ (reflected) to $H_y$ (incident) on the RIS top-surface. Due to the sub-wavelength structures that characterize the surface, the reflected near-field cannot be described in terms of a single plane wave. Thus, while numerical reflection coefficients can be computed, they do not have the same physical meaning as in the case of reflection from a homogeneous surface.

\section{Conclusion}

In this work, we have shown that the misconception of reciprocity breaking arises due to ignorance of mutual coupling and finite surface effects. Numerical reflection coefficients for a finite size RIS are computed and compared with the ideal coefficients as illustration. We use full-wave simulations and apply the electromagnetic reciprocity theorem to demonstrate that reciprocity is obeyed by an RIS even if its reflection phases are incident angle dependent. While the numerical study pertains to a particular RIS design, the Lorentz reciprocity theorem guarantees this outcome for any RIS free of nonlinear, active, or magneto-optic elements.

\appendices
\section{Technical Details of Full-wave Simulations}
\textit{Unit cell specifications}: With reference to the unit cell in Fig.~\ref{unitcellfig}, the dimensions are (in mm): $a_1=7.49$, $a_2=2.75$, $a_3=1.41$, $a_4 = 0.4$, $D=8.25$, $h=1.52$, with a Rogers RO4003C substrate (relative permittivity $\epsilon_r=3.55$ and loss tangent $\tan\delta=0.00242$ at 6.5 GHz ($\lambda = 46.12$ mm)).

\textit{Horn antenna specifications}: With reference to the identical horn antennas in Fig.~\ref{figfullsetup}, the dimensions are (in mm): $a=18.45$, $b=27.67$ (waveguide cross-section for single mode operation), $l=46.12$ (waveguide length), $d=l$ (flare length along axis), and $s=2$ ( scale factor between the horn aperture and waveguide cross-section dimensions).

\textit{Finite RIS specifications}: The horn antennas are placed at locations $A,B$ at angular locations $\theta=-65^{\circ},15^{\circ}$, respectively (clockwise angles w.r.t.~$+$z axis), with the antenna centers at an equal radial distance of $10\lambda=0.46122$ m from the center of the RIS top surface. For applying the FE-BI method (see Section \ref{sec3}), the antennas and RIS are enclosed in rectangular bounding boxes with an air gap of $\frac{\lambda}{10}=4.6$ mm in each dimension. To compute the local reflection coefficients of Fig.~\ref{refcoef}, two separate simulations are run in Comsol Multiphysics, the first without the RIS, and the second with the RIS. $H_y$ fields (corresponding to transverse magnetic polarization excitation) on the top surface of the RIS are computed in each case, which are processed to obtain the actual reflection coefficients.

\bibliographystyle{IEEEtran}
\bibliography{IEEEabrv,refs}

\begin{thebibliography}{10}
\providecommand{\url}[1]{#1}
\csname url@samestyle\endcsname
\providecommand{\newblock}{\relax}
\providecommand{\bibinfo}[2]{#2}
\providecommand{\BIBentrySTDinterwordspacing}{\spaceskip=0pt\relax}
\providecommand{\BIBentryALTinterwordstretchfactor}{4}
\providecommand{\BIBentryALTinterwordspacing}{\spaceskip=\fontdimen2\font plus
\BIBentryALTinterwordstretchfactor\fontdimen3\font minus
  \fontdimen4\font\relax}
\providecommand{\BIBforeignlanguage}[2]{{%
\expandafter\ifx\csname l@#1\endcsname\relax
\typeout{** WARNING: IEEEtran.bst: No hyphenation pattern has been}%
\typeout{** loaded for the language `#1'. Using the pattern for}%
\typeout{** the default language instead.}%
\else
\language=\csname l@#1\endcsname
\fi
#2}}
\providecommand{\BIBdecl}{\relax}
\BIBdecl

\bibitem{6g_a_research_direction}
C.~Pan, H.~Ren, K.~Wang, J.~F. Kolb, M.~Elkashlan, M.~Chen, M.~Di~Renzo,
  Y.~Hao, J.~Wang, A.~L. Swindlehurst, X.~You, and L.~Hanzo, ``Reconfigurable
  intelligent surfaces for {6G} systems: Principles, applications, and research
  directions,'' \emph{IEEE Commun. Mag.}, vol.~59, no.~6, pp. 14--20, Jun.
  2021.

\bibitem{zhang_dual_pol_2022}
N.~Zhang, K.~Chen, J.~Zhao, Q.~Hu, K.~Tang, J.~Zhao, T.~Jiang, and Y.~Feng, ``A
  dual-polarized reconfigurable reflectarray antenna based on dual-channel
  programmable metasurface,'' \emph{{IEEE} Trans. Antennas Propag.}, vol.~70,
  no.~9, pp. 7403--7412, Sep. 2022.

\bibitem{Sarris_propag_modeling_RIS}
Y.~Liu, Z.~Liu, S.~V. Hum, and C.~D. Sarris, ``An equivalence principle-based
  hybrid method for propagation modeling in radio environments with
  reconfigurable intelligent surfaces,'' \emph{{IEEE} Trans. Antennas Propag.},
  vol.~72, no.~7, pp. 5961--5973, Jul. 2024.

\bibitem{Di_Renzo_Smart_Radio_Environments_2020}
M.~D. Renzo, A.~Zappone, M.~Debbah, M.~S. Alouini, C.~Yuen, J.~D. Rosny, and
  S.~Tretyakov, ``Smart radio environments empowered by reconfigurable
  intelligent surfaces: How it works, state of research, and the road ahead,''
  \emph{{IEEE} J. Sel. Areas Commun.}, vol.~38, no.~11, pp. 2450--2525, Nov.
  2020.

\bibitem{Emara_reconfigurable_metasurface_varactor_2024}
M.~K. Emara, D.~Kundu, K.~Macdonell, L.~Rufail, and S.~Gupta, ``Reconfigurable
  metasurface reflectors using split-ring resonators with co-designed biasing
  for magnitude/phase control,'' \emph{{IEEE} Trans. Antennas Propag.},
  vol.~72, no.~9, pp. 7425--7430, Sep. 2024.

\bibitem{chen_angle-dependent_2020}
W.~Chen, L.~Bai, W.~Tang, S.~Jin, W.~X. Jiang, and T.~J. Cui, ``Angle-dependent
  phase shifter model for reconfigurable intelligent surfaces: Does the
  angle-reciprocity hold?'' \emph{{IEEE} Commun. Lett.}, vol.~24, no.~9, pp.
  2060--2064, Sep. 2020.

\bibitem{Yue_reciprocity_tvt_2023}
S.~Yue, S.~Zeng, H.~Zhang, F.~Lin, L.~Liu, and B.~Di, ``Intelligent
  omni-surfaces aided wireless communications: {Does} the reciprocity hold?''
  \emph{IEEE Trans. Veh. Technol.}, vol.~72, no.~6, pp. 8181--8185, Jun. 2023.

\bibitem{10360391}
W.~Tang, J.~Wang, J.~Y. Dai, M.~Di~Renzo, S.~Jin, Q.~Cheng, and T.~J. Cui, ``On
  path loss and channel reciprocity of {RIS}-assisted wireless
  communications,'' in \emph{Intelligent Surfaces Empowered {6G} Wireless
  Network}.\hskip 1em plus 0.5em minus 0.4em\relax Wiley-IEEE Press, 2024, pp.
  37--58.

\bibitem{yu_light_propag_2011}
N.~Yu, P.~Genevet, M.~A. Kats, F.~Aieta, J.-P. Tetienne, F.~Capasso, and
  Z.~Gaburro, ``Light propagation with phase discontinuities: Generalized laws
  of reflection and refraction,'' \emph{Science}, vol. 334, no. 6054, pp.
  333--337, 2011.

\bibitem{cui_3_bit_angle_insensitive_2022}
J.~C. Liang, Q.~Cheng, Y.~Gao, C.~Xiao, S.~Gao, L.~Zhang, S.~Jin, and T.~J.
  Cui, ``An angle-insensitive 3-bit reconfigurable intelligent surface,''
  \emph{{IEEE} Trans. Antennas Propag.}, vol.~70, no.~10, pp. 8798--8808, Oct.
  2022.

\bibitem{liang_low_angular_sensitivity_2024}
J.~C. Liang, Y.~Gao, Z.~W. Cheng, R.~Z. Jiang, J.~Y. Dai, L.~Zhang, Q.~Cheng,
  S.~Jin, and T.~J. Cui, ``An optically transparent reconfigurable intelligent
  surface with low angular sensitivity,'' \emph{Adv. Opt. Mater.}, vol.~12,
  no.~6, p. 2202081, 2024.

\bibitem{zhao_angular_stability_for_precise_wave_manipulation_2025}
W.~Zhao, S.~Wang, K.~Tang, K.~Chen, T.~Jiang, J.~Zhao, and Y.~Feng,
  ``Hybrid-modulated programmable metasurface with angular stability for
  precise electromagnetic wave manipulation,'' \emph{Laser Photonics Rev.}, p.
  e01115, 2025.

\bibitem{chewemt}
W.~C. Chew, ``Lectures on electromagnetic field theory,''
  \url{https://engineering.purdue.edu/wcchew/ece604f20/EMFTAll.pdf}, 2015,
  accessed: Aug 07, 2025.

\bibitem{balanis2016antenna}
C.~A. Balanis, \emph{Antenna Theory: Analysis and Design}.\hskip 1em plus 0.5em
  minus 0.4em\relax Hoboken, NJ, USA: John Wiley \& Sons, 2016.

\bibitem{sai}
S.~Sanjay~Narayanan, U.~K. Khankhoje, and R.~Krishna~Ganti, ``Optimum
  beamforming and grating-lobe mitigation for intelligent reflecting
  surfaces,'' \emph{{IEEE} Trans. Antennas Propag.}, vol.~72, no.~11, pp.
  8540--8553, Nov. 2024.

\bibitem{harrington1961time}
R.~F. Harrington, \emph{Time-Harmonic Electromagnetic Fields}, ser. McGraw-Hill
  Texts in Electrical Engineering.\hskip 1em plus 0.5em minus 0.4em\relax New
  York, NY, USA: McGraw-Hill, 1961.

\bibitem{Abrardo2024}
A.~Abrardo, A.~Toccafondi, and M.~Di~Renzo, ``Design of reconfigurable
  intelligent surfaces by using {S}-parameter multiport network
  theory—optimization and full-wave validation,'' \emph{{IEEE} Trans.
  Wireless Commun.}, vol.~23, no.~11, pp. 17\,084--17\,102, 2024.

\end{thebibliography}

\end{document}